\documentclass[letter, twocolumn]{jpsj3} 

\def\runauthor{Online-Journal Subcommittee}

\title{%
Frustration-Induced Ferrimagnetism in $S=1/2$ Heisenberg Spin Chain \\
}

\author{%
Tokuro Shimokawa\thanks{E-mail address: rk09s002@stkt.u-hyogo.ac.jp }
 and Hiroki Nakano\thanks{E-mail address: hnakano@sci.u-hyogo.ac.jp}
}

\inst{%
Graduate School of Material Science, University of Hyogo, Kamigori, Hyogo 678-1297, Japan
}

\recdate{\today}

\abst{%
The ground-state properties 
of the $S=1/2$ frustrated Heisenberg spin chain 
with interactions up to fourth nearest neighbors are investigated 
by the exact-diagonalization  method and 
density matrix renormalization group method.
Our numerical calculations clarify that 
the ferrimagnetic state is realized 
in the ground state 
in spite of the fact that a multi-sublattice structure 
in the shape of the system is absent. 
We find that there are two types of ferrimagnetic phases: 
one is the well-known ferrimagnetic phase 
of the Lieb-Mattis type and 
the other is the nontrivial ferrimagnetic phase 
that is different from that of the  Lieb-Mattis type. 
Our results suggest that a multi-sublattice structure 
of the shape is not necessarily required for the occurrence 
of ferrimagnetism. 
}

\kword{%
quantum spin chain, frustration, ferrimagnetism, DMRG, exact diagonalization
}

\begin{document}
\maketitle

Ferrimagnetism is one of fundamental phenomena 
in the field of magnetism. 
A typical case showing ferrimagnetism 
is that when a system includes spins of two types 
that antiferromagnetically interact 
between two spins of different types in each neighboring pair. 
The simplest example is an ($S$, $s$)=($1$, $1/2$) 
antiferromagnetic mixed spin chain, in which  
two different spins are arranged alternately in a line 
and coupled by the nearest-neighbor 
antiferromagnetic interaction\cite{Sakai}. 
The occurrence of ferrimagnetism in this case 
is understood within the Marshall-Lieb-Mattis theorem 
concerning quantum spin systems\cite{Lieb, Marshall}. 
Even though a system includes spins of one type, 
this theorem also derives the presence of ferrimagnetism 
when the system includes 
more than one sublattice of spin sites, 
for example, the spin system in a diamond chain
\cite{Takano, Okamoto, Tonegawa, SpinGap, comments_diamond}. 
From these two mechanisms, the existence 
of a multi-sublattice structure is very important 
for the occurrence of ferrimagnetism.  

At this stage, one asks a fundamental question: 
Is a multi-sublattice structure in the shape 
of a Hamiltonian essential and necessary 
for the occurrence of ferrimagnetism? 
The purpose of the present study is to answer this question. 
Our following demonstration will clarify that 
the answer is no. 
In this study, we find that ferrimagnetism can appear 
due to the effect of magnetic frustration 
even in the absence of a multi-sublattice structure 
in the shape of a system. 

In this study, we examine the model whose Hamiltonian is
given by 
\begin{eqnarray}
\label{Hamiltonian}
\mathcal{H} &=& 
J \sum_{i}  [{\bf S}_{i}\cdot {\bf S}_{i+1}  
+  \frac{1}{2} {\bf S}_{i}\cdot {\bf S}_{i+2}] \\ \nonumber
            &-& 
J^{\prime} \sum_{i}  [{\bf S}_{i}\cdot {\bf S}_{i+3} 
+ \frac{1}{2} ( {\bf S}_{i}\cdot {\bf S}_{i+2} 
+  {\bf S}_{i}\cdot {\bf S}_{i+4}) ],
\end{eqnarray}
where ${\bf S}_{i}$ is the $S=1/2$ spin operator at the site $i$.
The system size is denoted by $N$. 
We emphasize here that this model has 
only one spin in a unit cell, namely, 
it has no sublattice structure. 
Energies are measured in units of $J$; 
therefore, we set $J=1$ hereafter. 
We have a controllable parameter, $J^{\prime}$, 
in the Hamiltonian (\ref{Hamiltonian}). 
This model was originally introduced in ref.~\ref{Nakano} 
detailing the study of constructing a model Hamiltonian 
as a generalization from the Majumdar-Ghosh model\cite{Majumdar}. 
The Hamiltonian (\ref{Hamiltonian}) includes two cases 
in which the ground state of the system is exactly obtained. 
For $J^{\prime}=0$, the system is reduced to 
the Majumdar-Ghosh model\cite{Majumdar}, 
whose ground state is described by direct products of
spin-singlet states in nearest-neighbor pairs 
of $S=1/2$ spins. 
The ground state is called the dimer (DM) state. 
Note that even if $J^{\prime}$ takes a nonzero value, 
this DM state is still an eigenstate of the system. 
The DM state becomes an excited state 
when $J^{\prime}$ increases.
In the limit of a large $J^{\prime}$, on the other hand, 
the ferromagnetic (FM) state becomes the ground state.  
Although the wavefunctions of these limits are well known, 
the ground state in the intermediate region 
is not sufficiently understood. 
In ref. \ref{Nakano}, it was reported that 
the spontaneous magnetization in the intermediate region 
appears and that the magnetization changes gradually. 
In the present study, we investigate the magnetic structure 
of the ground state in this intermediate region 
by some numerical calculations.
We show that our results lead to the conclusion 
that the ferrimagnetic state can appear in the ground state, 
even of models consisting of only a spin in each unit cell.

We employ two reliable numerical methods, 
the exact diagonalization (ED) method and 
density matrix renormalization group (DMRG) 
method\cite{DMRG1,DMRG2}. 
The ED method can be used to obtain 
precise physical quantities for finite-size clusters.  
This method does not suffer from the limitation 
of the shape of the clusters. 
It is applicable even to systems with frustration, 
in contrast to the quantum Monte Carlo (QMC) method 
coming across the so-called negative-sign problem 
for a system with frustration. 
The disadvantage of the ED method is 
the limitation that the available sizes are only small. 
Thus, we should pay careful attention to 
finite-size effects in quantities obtained from this method. 
On the other hand, the DMRG method is very powerful 
when a system is one-dimensional 
under the open-boundary condition. 
The method can treat much larger systems than the ED method 
and is applicable even to a frustrated system. 
In the present research, we use the "finite-system" DMRG method. 

In the present study, 
two quantities are calculated by the two methods 
mentioned above. 
One is the lowest energy in each subspace divided 
by $S_{z}^{\rm tot}$ to determine 
the spontaneous magnetization $M$, 
where $S_{z}^{\rm tot}$ is the $z$ component of the total spin. 
We can obtain the lowest energy $E(N,S_{z}^{\rm tot},J^{\prime})$ 
for a system size $N$ and a given $J^{\prime}$.  
For example, the energies of each $S_{z}^{\rm tot}$ 
in the three cases of $J^{\prime}$ are presented in Fig. 1(a). 
This figure is obtained by our DMRG calculations 
of the system of $N=96$ with the maximum number of retained states
 ($MS$) 1500, and a number of sweeps ($SW$) 50. 
The spontaneous magnetization $M$ 
for a given $J^{\prime}$ is determined 
as the highest $S_{z}^{\rm tot}$ 
among those at the lowest common energy.  
(See arrows in Fig. 1(a).) 
The other quantity is the local magnetization in the ground state 
for investigating the spin structure of the state. 
The local magnetization is obtained 
by calculating $\langle S_{i}^{z} \rangle$, 
where $\langle A \rangle$ denotes the expectation value 
of the physical quantity $A$ and 
$S_{i}^{z}$ is the $z$-component of the spin at the site $i$.

\begin{figure}[t]
\begin{center}
\includegraphics[width=5.5cm]{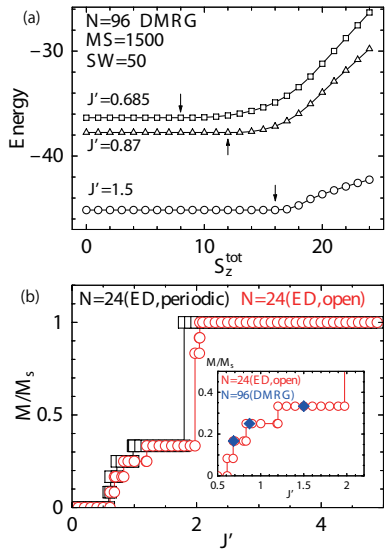}
\caption{(Color) (a) Lowest energy in each subspace divided 
by $S_{z}^{\rm tot}$. 
Results of the DMRG calculations are presented 
when the size system is $N=96$ 
for $J^{\prime}=0.685,  0.87$, and 1.5. 
Arrows indicate the spontaneous magnetization $M$ 
for a given $J^{\prime}$; $M$ is determined 
to be the highest $S_{z}^{\rm tot}$ 
among the values taking the lowest common energy. 
(b) $J^{\prime}$ dependence 
of the normalized magnetization $M/M_{\rm s}$ 
in the ground state. 
Red circles (black squares) denote the results obtained 
by ED calculations for a size system of $N=24$ 
under the open (periodic)-boundary condition. 
In the inset of (b), blue diamonds show 
the results obtained by DMRG calculations 
for a size system of $N=96$ under the open-boundary condition 
accompanied by red circles denoting 
the results obtained by ED calculations for a size system of $N=24$ 
under the open-boundary condition. 
}
\label{fig1}
\end{center}
\end{figure}

\begin{figure}[ht]
\begin{center}
\includegraphics[width=3.7cm, angle=90]{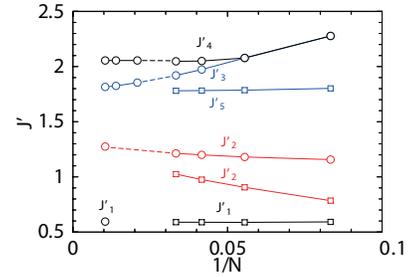}
\caption{(Color) Size dependences of the phase boundaries. 
The results presented are those of $N=12, 18, 24$, and 30 from 
the ED calculations 
and those of $N=48, 72$, and 96 from the DMRG calculations. 
Squares (circles) denote results in the cases 
under the periodic (open)-boundary condition. 
Dotted lines are drawn as guides for the eyes  
between the data from the ED and DMRG calculations. 
In the limit $N \rightarrow \infty$, 
the phase boundary 
between the $0<M/M_{\rm s} < 1/3$ phase and the $M/M_{\rm s}=1/3$ phase 
seems to converge to approximately $J^{\prime} = 1.30$.}
\label{fig2}
\end{center}
\end{figure}

First, let us examine 
the $J^{\prime}$ dependence of $M/M_{\rm s}$ 
to confirm the existence of the intermediate phase 
between the FM phase and the nonmagnetic DM phase 
irrespective of the boundary conditions, 
where $M_{\rm s}$ is the saturation value of the magnetization.
Results are presented for $N=24$ from our ED calculations 
under the open and periodic boundary conditions 
in Fig. \ref{fig1}(b). 
We successfully observe the intermediate-magnetization phase 
irrespective of the boundary conditions. 
We also include in Fig. \ref{fig1}(b) some DMRG results of $N=96$,  
which suggests a weak size dependence of $M/M_{\rm s}$ 
as a function of $J^{\prime}$.  
Careful observation of the region of $0<M/M_{\rm s}\leq 1/3$ 
enables us to find that the intermediate-magnetization phase 
consists of two phases. 
One is the phase where $M/M_{\rm s}$ is fixed at $1/3$; 
this feature is that of the ferrimagnetism of 
the so-called Lieb-Mattis (LM) type, 
in which the spontaneous magnetization is fixed 
to be a simple fraction 
of the saturated magnetization\cite{Lieb, Marshall}. 
The other is the phase where $M/M_{\rm s}$ changes continuously 
with respect to the strength of $J^{\prime}$. 
This feature is certainly different from that 
of the LM ferrimagnetism; 
the continuous change in $M/M_{\rm s}$ is observed 
as the ferrimagnetism of the non-Lieb-Mattis (NLM) type 
in several models\cite{PF1, PF2, PF3, PF4, PF5, PF6, PF7, 
Nakano-2D}. 
We will determine later whether or not the phase 
of $0<M/M_{\rm s}<1/3$ in the present model is 
of the NLM type. 
Note here that these two phases are observed 
under both boundary conditions. 
On the other hand, the region of $1/3<M/M_{\rm s}<1$ is observed 
near $M/M_{\rm s}=1$ only under the open-boundary condition. 
At present, it is unclear whether or not this phase survives 
in the limit $N \rightarrow \infty$. 

Next, we study 
the size dependences of the boundaries between the phases 
observed above. 
We investigate five boundaries: 
$J^{\prime}=J_{1}^{\prime}$ 
between the DM phase and the phase of $0<M/M_{\rm s}<1/3$,  
$J^{\prime}=J_{2}^{\prime}$
between the phase of $0<M/M_{\rm s}<1/3$ and the phase 
of $M/M_{\rm s}=1/3$,  
$J^{\prime}=J_{3}^{\prime}$
between the phase of $M/M_{\rm s}=1/3$ and 
the phase of $1/3<M/M_{\rm s}<1$,  
$J^{\prime}=J_{4}^{\prime}$ 
between the phase of $1/3<M/M_{\rm s}<1$ and the FM phase, 
and 
$J^{\prime}=J_{5}^{\prime}$ 
between the phase of $M/M_{\rm s}=1/3$ and the FM phase 
without the phase of $1/3<M/M_{\rm s}<1$. 
Note that $J^{\prime}_{3}$ and $J^{\prime}_{4}$ appear 
under the open-boundary condition, whereas 
$J^{\prime}_{5}$ appears 
under the periodic-boundary condition. 
Figure \ref{fig2} shows the results 
of $N=12, 18, 24$, and 30 from the ED calculations 
and those of $N=48, 72$, and 96 from the DMRG calculations. 
One finds that $J^{\prime}_{1}$ from the ED calculations 
under the periodic-boundary condition and 
that from the DMRG calculations under the open-boundary condition 
are consistent with each other; we have 
$J^{\prime}_{1}\sim 0.59$ as an extrapolated value. 
Concerning the boundary $J^{\prime}_{2}$, 
there exists a not so small difference 
between the result under the open-boundary condition 
and that under the periodic-boundary condition 
for a given $N$; however, 
$J^{\prime}_{2}$ seems to converge to 1.30 
irrespective of the boundary condition. 
On the other hand, the situations of the boundaries 
of the phase of $M/M_{\rm s}=1/3$ and the FM phase 
are slightly complicated in our results. 
It seems that $J^{\prime}_{3}$ and $J^{\prime}_{4}$ 
become farther away from each other with increasing $N$ 
and 
that $J^{\prime}_{3}$ and $J^{\prime}_{5}$ converge 
to the same value of 1.77. 
We also have $J^{\prime}_{4}$ converging to 2.06. 
From these results of the extrapolation, 
it is evident that the phase of $M/M_{\rm s}=1/3$ and 
the phase of $0<M/M_{\rm s}<1/3$ exist 
in the thermodynamic limit. 
On the other hand, it is difficult to determine
whether or not the phase of $1/3<M/M_{\rm s}<1$ is present. 
There is a possibility 
that this phase merges with the FM phase in the thermodynamic limit 
for two reasons: 
one is that this phase appears only near $M/M_{\rm s}=1$ 
and the other is that it is observed only 
under the open-boundary condition. 
The issue of whether or not this phase survives should be 
clarified in future studies; 
hereafter, we do not pay further attention to this phase. 

\begin{figure}[t]
\begin{center}
\includegraphics[width=4.5cm]{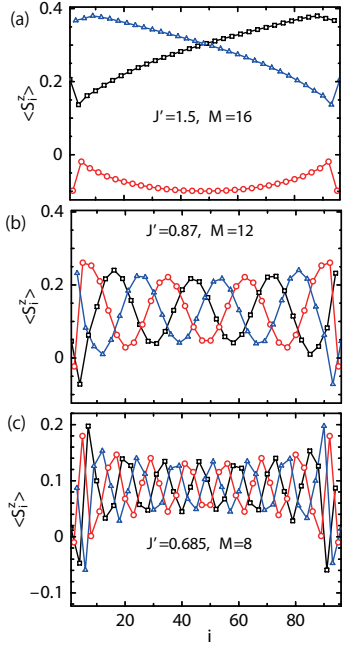}
\caption{(Color) Local magnetization $\langle S_{i}^{z} \rangle$ 
under the open-boundary condition: 
(a) for $J^{\prime}=1.5$, (b) for $J^{\prime}=0.87$, 
and (c) for $J^{\prime}=0.685$ 
from the DMRG calculation for $N=96$. 
The site number is denoted by $i$, which is classified 
into $i=3n-2$, $3n-1$, and $3n$, where $n$ is an integer.
Squares, circles, and triangles mean $i=3n-2$, $3n-1$, and $3n$, 
respectively. }
\label{fig3}
\end{center}
\end{figure}

Next, we examine the local magnetization $\langle S_{i}^{z} \rangle$ 
in the two phases of $0<M/M_{\rm s}<1/3$ and $M/M_{\rm s}=1/3$ 
to determine the magnetic properties in each phase. 
We present our DMRG results of $\langle S_{i}^{z} \rangle$ 
of the system of $N=96$. 
Note here that we calculate $\langle S_{i}^{z} \rangle$ 
within the subspace of the highest $S_{z}^{\rm tot}$ 
corresponding to the spontaneous magnetization $M$ obtained 
for a given $J^{\prime}$. 
The results of $\langle S_{i}^{z} \rangle$ are shown 
in Figs. \ref{fig3}(a)-3(c) 
for $J^{\prime}=0.685$, 0.87, and 1.5, respectively. 
In each case, one can observe a three-sublattice structure 
of the spin state clearly. 
In Fig. \ref{fig3}(a), 
the $i$ dependence of $\langle S_{i}^{z} \rangle$ 
in each of the sublattices of the spin structure  
is weak around the center of the system, 
although the edge effect spreads into a wide range 
from the edges. 
This behavior suggests that the spin state forms 
the LM ferrimagnetic state of up-up-down, 
which is consistent with $M/M_{\rm s}$=1/3 
in the parameter region near approximately $J^{\prime}=1.5$. 
In Figs. \ref{fig3}(b) and 3(c), on the other hand, we find that 
the local magnetization shows 
a longer-distance periodicity in addition to the 
three-sublattice structure. 
The longer-distance periodicity changes when $J^{\prime}$ is 
changed within the phase of $0<M/M_{\rm s}<1/3$, the periodicity 
suggests an incommensurate modulation. 
A similar feature of this local structure was reported 
in some one-dimensional quantum frustrated spin 
systems\cite{PF4, PF5}. 
Therefore, the phase of $0<M/M_{\rm s}<1/3$ is considered 
as the NLM-type ferrimagnetic phase. 
This incommensurate feature originates 
from the effects of quantum fluctuation and frustration. 
We also calculate $\langle S_{i}^{z} \rangle$ 
for different system sizes, $N=48$ and 72. 
At least from these data (not shown in this papar), 
the periodicity and amplitude of the modulation 
seem to show only weak dependences on the system size. 
Note that the behavior of long-distance periodicity 
accompanied by the three-sublattice structure at the same time 
is different from the wave functions with a long periodicity 
reported in ref.~\ref{Schulenburg_Richter}. 

\begin{figure}[t]
\begin{center}
\includegraphics[width=5.5cm]{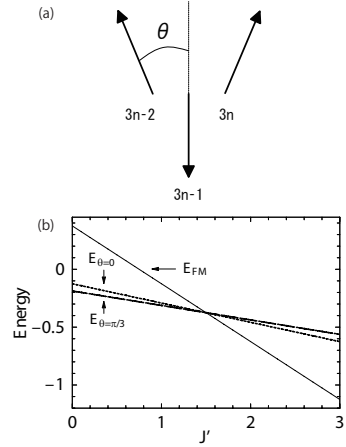}
\caption{(a) Spin configuration from the point of view 
of classical vectors. 
The site number $i$ in the Hamiltonian (\ref{Hamiltonian}) 
is classified into $3n$, $3n-1$, and $3n-2$, 
where $n$ is a positive integer. 
The angle $\theta$ for $J^{\prime}$ 
is determined by minimizing the classical energy. 
(b) $J^{\prime}$ dependences of classical energies 
of eq. (2) for $\theta=0$ (LM, dotted line), 
$\pi/3$ (NM, dotted chain line), and 
FM (solid line) energy of eq. (3).}
\label{fig4}
\end{center}
\end{figure}

Here, let us discuss the behavior of the intermediate 
phase between the FM phase and the nonmagnetic phase 
from the viewpoint 
that spins in the Hamiltonian (\ref{Hamiltonian}) 
are assumed to be classical vectors. 
We consider the spin configuration of the classical vectors 
depicted in Fig. \ref{fig4}(a),
where the characteristic angle $\theta$ is defined. 
This classical spin arrangement has been determined 
from our observation in Fig. 3 
that the three-sublattice spin structure 
is realized in the intermediate region.  
The case of $\theta=0$ means 
that this classical state is the LM-type ferrimagnetic state 
with the ratio of the spontaneous magnetization to the 
saturated magnetization to be $1/3$. 
On the other hand, $\theta=\pi/3$ means 
that the state is in a nonmagnetic (NM) state. 
The classical energy per spin site 
under the periodic-boundary condition is given by
\begin{equation}
E(J^{\prime}, \theta) =  
\frac{1}{24}[(6-4J^{\prime}){\rm cos}^{2} \theta -(6-4J^{\prime}) {\rm cos} \theta +(-3-4J^{\prime})],
\label{ene_classic_ferri}
\end{equation}
and the energy of the ferromagnetic state is given by
\begin{equation}
E_{\rm FM}=(3-4J^{\prime})/8.
\label{ene_classic_fm}
\end{equation}
The dependences of the energies shown in eqs. (\ref{ene_classic_ferri}) 
and (\ref{ene_classic_fm}) are shown in Fig. \ref{fig4}(b). 
The FM (NM) phase appears at $J^{\prime}>1.5$ ($J^{\prime}<1.5$). 
One finds that $J^{\prime}$=1.5 is the boundary 
of the FM and NM phases. 
At exactly $J^{\prime}$=1.5, many states degenerate, 
including not only  the FM and NM states 
but also the ferrimagnetic state 
with an arbitrary angle $\theta$. 
There is no intermediate phase between the two phases. 
It is worth emphasizing here that even the LM ferrimagnetic phase 
does not appear. 
This arguement suggests that the occurrence 
of the intermediate-magnetization state 
observed in the Hamiltonian (\ref{Hamiltonian}) 
of the quantum system is a consequence of 
the quantum effect induced by frustration. 

Finally, we mention another case 
when the intermediate-magnetization phase 
appears in the frustrated spin system in one dimension 
with anisotropic interactions\cite{Tonegawa1,Tonegawa2,
Tonegawa3,Tonegawa4,Tonegawa5}. 
Note here that this phase disappears 
in the isotropic case of interactions, which suggests that 
the origin of this phase is the anisotropy. 
However, it has not been examined yet whether or not
this model shows a similar incommensurate modulation. 
Such examination would clarify the relationship 
between the intermediate magnetization of this model and 
the NLM ferrimagnetism studied in the present case. 


In summary, 
we study the ground-state properties 
of an $S=1/2$ frustrated Heisenberg spin chain 
with isotropic interactions up to the fourth nearest neighbor 
by the ED and DMRG methods.
In spite of the fact that this system consists of 
only a single spin site in each unit cell determined 
from the shape of the Hamiltonian, 
the ferrimagnetic ground state is surprisingly realized 
in a finite region 
between the ferromagnetic and nonmagnetic states. 
This result is in contrast to that of other systems 
of translationally invariant chains\cite{uni_chain1,uni_chain2}.
We find that the intermediate region consists of phases 
of two ferrimagnetic types, 
the Lieb-Mattis type and non-Lieb-Mattis type. 
In the latter phase, we confirm that 
the local magnetization shows characteristic 
incommensurate modulation. 
The presence of the ferrimagnetic state 
without a sublattice structure of the shape of the system 
is a consequence of the strong quantum effect induced 
by frustration. 
Our findings shed light on a new aspect 
of the effect of frustration in quantum systems. 

\paragraph{Acknowledgments}\

We wish to thank Prof.~K.~Hida and Prof.~T.~Tonegawa 
for fruitful discussions. 
This work was partly supported  
by a Grant-in-Aid (No.20340096) 
from the Ministry of Education, Culture, Sports, 
Science and Technology of Japan.
This work was partly supported by 
a Grant-in-Aid (No. 22014012) 
for Scientific Research and Priority Areas 
``Novel States of Matter Induced by Frustration'' 
from the Ministry of Education, Culture, Sports, Science 
and Technology of Japan. 
Diagonalization calculations in the present work were 
carried out based on TITPACK Version 2 coded by H. Nishimori.
DMRG calculations were carried out 
using the ALPS DMRG application\cite{ALPS}.
Some of the calculations were carried out 
at the Supercomputer Center, Institute for Solid State Physics, 
University of Tokyo.


\begin{thebibliography}{99}
\bibitem{comments_diamond}
As a candidate compound of the diamond chain system, 
natural mineral azurite, ${\rm Cu_{3}(CO_{3})_{2}(OH)_{2}}$, 
is proposed\cite{Kikuchi, Ohta}.
\end{thebibliography}
\end{document}